\documentclass[showpacs,amsmath,amssymb,floatfix,prd,superscriptaddress,footinbib,twocolumn]{revtex4-1}
\usepackage{amssymb}
\usepackage{graphics}
\usepackage{epstopdf}
\usepackage{epsfig}
\usepackage{dcolumn}
\usepackage{bm}
\renewcommand{\vec}[1]{\mathbf{#1}}
 
\newcommand{\abs}[1]{\left| #1 \right|} 
\renewcommand{\d}[2]{\frac{d #1}{d #2}} 
\newcommand{\pd}[2]{\frac{\partial #1}{\partial #2}} 
 
\let\baraccent=\= 
\renewcommand{\=}[1]{\stackrel{#1}{=}} 

\usepackage{color}
 \definecolor{blue}{rgb}{0,0,1}
 \definecolor{sepia}{rgb}{0,0.8,0.2}

\newcommand{\circone}{\mbox{\textcircled{\tiny 1}}}%
\newcommand{\circtwo}{\mbox{\textcircled{\tiny 2}}}%
\newcommand{\circx}{\mbox{\textcircled{\footnotesize x}}}%

\begin{document}

\title{Classical scattering of charged particles confined on an inhomogeneous helix}



\author{A. V.  Zampetaki}
\author{J. Stockhofe}
\author{S. Kr\"{o}nke}
\affiliation{Zentrum f\"{u}r Optische Quantentechnologien, Universit\"{a}t Hamburg, Luruper Chaussee 149, 22761 Hamburg, Germany}
\author{P. Schmelcher}
\affiliation{Zentrum f\"{u}r Optische Quantentechnologien, Universit\"{a}t Hamburg, Luruper Chaussee 149, 22761 Hamburg, Germany}
\affiliation{The Hamburg Centre for Ultrafast Imaging, Luruper Chaussee 149, 22761 Hamburg, Germany}
\date{\today}

\begin{abstract}
We explore the effects arising due to the coupling of the center of mass and relative motion of two charged particles confined on an inhomogeneous helix with a 
locally modified radius. It is first proven that a separation of the center of mass and the relative motion is provided if and only if the confining manifold represents a homogeneous helix.
 In this case
 bound states of repulsively Coulomb interacting particles occur. For an inhomogeneous helix, the coupling of the center of mass 
and relative motion induces an 
energy transfer between the collective and relative motion, leading to dissociation of initially bound states in a scattering process. 
Due to the time reversal symmetry, a binding of the particles out of the scattering continuum is thus equally possible.
 We identify the regimes of dissociation for different initial conditions
 and provide an analysis of the underlying phase space via Poincar\'{e} surfaces of section. Bound states inside the inhomogeneity as well as  resonant states are identified.
\end{abstract}

\pacs{37.10.Ty, 37.90.+j, 05.45.-a, 45.50.-j}
\maketitle

\begin{center}
  {\bf{I. INTRODUCTION}} 
\end{center}

The formation of helical patterns and structures is common in many natural systems  ranging from  DNA molecules and amino-acids to non-neutral plasmas trapped in  magnetic fields \cite{Totsuji88} and
self-assembled  configurations of charged particles confined in nanotubes \cite{Vernizzi09}. Studying the motion of particles confined in a helix has proven
 to be a useful tool for the understanding of complex phenomena such as the optical activity of sugar solutions 
\cite{Desobry73,Dufey06}. Certainly the problem of the confined motion of particles in a helical manifold is of fundamental interest since it reveals many intriguing phenomena.
Quantum particles confined in one dimension (1D), preserve some information of the surrounding 3D space and thus experience an 
effective geometric potential which depends on the curvature of the confining manifold \cite{Costa81}. Such geometric potential effects lead to the formation of bound states 
 in helical waveguides with a locally modified radius \cite{Exner07}
or in twisting tubes \cite{Goldstone92}. In the presence of an electric field, super-lattice properties can emerge for a confined charge carrier \cite{Kibis05},
whereas when the particles interact with dipolar forces a peculiar quantum phase transition from liquid to gas has been predicted \cite{Law08}.

In spite of their physical interest, helical traps have only recently been  investigated experimentally. In nano\-tech\-nology, 
curved nanotubes such as rolls, spirals and helices from thin solid films of silicon-germanium can be constructed \cite{Prinz00,Schmidt01}.
Helical traps can also be realized experimentally for cold atoms either via the interference
 of counter-propagating Laguerre-Gaussian beams \cite{Bhatt07,Okulov12,Vargas10} or via the evanescent field  of a nanofiber \cite{Vetsch10,Vetsch12,Reitz12} which creates a double-helix  trapping potential.
Such setups allow  the creation of a homogeneous  
helical potential over 
the entire length of the nanofiber as well as local modifications of the radius and the pitch of the helix through local variations of the diameter of the nanofiber \cite{Reitz12}.
Beyond these, a plethora of trapping techniques  also exist for (ultra-) cold ions \cite{Paul90, Schneider10}.

Motivated by the above, it is instructive to explore the classical behaviour of ions or generally charged particles in a helical geometry. Surprisingly this problem 
has not been studied extensively in the literature. In ref. \cite{Schmelcher11}  it has been shown that the classical dynamics of a 
system of identical charged particles confined in a helical manifold presents very intriguing phenomena when the particles interact via long-range 
interactions such as the Coulomb interaction.
 In particular, the interplay between the 1D confined  motion of the particles
and their interactions via the full 3D space gives rise to an effective oscillatory force. This fact yields, in turn, stable equilibrium configurations despite 
the repulsive interactions between the particles and induces classical bound states whose number can be tuned by varying the parameters of the helix.

Following the direction of the above study, we explore in the present work the two-body scattering dynamics off an inhomogeneity in a helical trap. As a first step 
we rigorously prove that a separation of the center of mass (CM) and the relative motion is provided  for an interaction potential which depends exclusively on the Euclidean distance 
between the particles $V(\abs{\vec{r}_1-\vec{r}_2})$, if and only if the confining curve is  a homogeneous helix. Then, we examine the case of an inhomogeneous helical trap with 
a locally modified radius,
and explore effects due to the coupling of the CM and relative motion. It is shown that initially bound states can finally 
dissociate due to the modulation of the potential which leads to an energy transfer between the CM and the relative degrees of freedom. Due to time reversal 
symmetry, it is thus equally possible for two unbound charged particles to form a bond due to the local inhomogeneity. A phase space analysis
provides us with bound states within the inhomogeneous region as well as with resonant states and completes the picture of the two-particle dynamics.

The paper is organized as follows. In Sec. II we present the general Lagrangian for the problem of two interacting classical charged  particles confined on a curve and we  investigate 
the properties that the confining curve  has to fulfill so that a separation of the CM and the relative degrees of freedom is provided. In Sec. III we present our model of two 
charged particles confined to an inhomogeneous helix. Section IV contains our results for the scattering, whereas section V 
provides our analysis of the respective phase space. Finally,  Sec. VI represents a brief summary of our findings.

\begin{center}
 { \bf{II. INTERACTING PARTICLES CONFINED TO A CURVED 1D-MANIFOLD}} 
\end{center}

We consider a system of two particles with masses $m_1, m_2$, interacting via a potential $V(\abs{\vec{r}_1-\vec{r}_2})$ that depends only on the  Euclidean distance between them. 
 Their Lagrangian is given by
\[L(\{\vec{r}_i, \dot{\vec{r}}_i\})=\frac{1}{2} \sum_{i=1}^2  m_i {\dot{\vec{r}}_i}^2- V(\abs{\vec{r}_1-\vec{r}_2}).\]
If the particles are confined onto a smooth, regular and either
closed or infinitely extended space curve $\vec{r}:\mathbb{R}\mapsto \mathbb{R}^3$
parametrized with the arbitrary parameter $u$, i.e. $\vec{r}_i=\vec{r}(u_i)$, the Lagrangian takes the 
form
\begin{eqnarray}
L(\{u_i, \dot{u}_i\})&=&\frac{1}{2} \sum_{i=1}^2  m_i \abs{\partial_{u_i}\vec{r}(u_i)}^2\dot{u}_i^2 \nonumber \\
&-& V(\abs{\vec{r}(u_1)-\vec{r}(u_2)}).\label{lanui}
\end{eqnarray}
 If we choose the arc  length parametrization \cite{footnote}
\begin{equation}
s:u\mapsto s(u)=\int_0^u \abs{\partial_{u'}\vec{r}(u')}du', \label{alp1}
\end{equation}
since  the tangent vector $\vec{t}(s_i) = \partial_{s_i}\vec {r} (s_i)$ is a unit vector we arrive at the expression
\begin{equation}
L(\{s_i, \dot{s}_i\})=\frac{1}{2} \sum_{i=1}^2  m_i \dot{s}_i^2 
- V(\abs{\vec{r}(s_1)-\vec{r}(s_2)}). \label{fueq1}
\end{equation}
We thus observe that the kinetic energy term retains the Cartesian form in the arc length parametrization, leading to the familiar expressions for the conjugate momenta and the
Euler-Lagrange (EL) equations of motion.
Introducing the CM $S=(m_1 s_1 +m_2 s_2)/(m_1+m_2)$ and the relative coordinate $s=s_1-s_2$,
 as well as the total mass $M=m_1+m_2$ and the reduced mass $\mu=\frac{m_1 m_2}{M}$, we are led to
\begin{equation}
L(\{s, S, \dot{s}, \dot{S}\})=\frac{1}{2} M \dot{S}^2+\frac{1}{2} \mu \dot{s}^2 
- \tilde{V}\left(S,s\right), \label{fueq3}
\end{equation}
where
\begin{equation}
 \tilde{V}(S,s)=V\left(\abs{\vec{r}(S+\frac{m_2}{M}s)-\vec{r}(S-\frac{m_1}{M}s)}\right). \label{potsep1}
\end{equation}

This yields the following EL equations

\begin{equation}
M\ddot{S} = - \pd{\tilde{V}}{S},~\mu \ddot{s} = - \pd{\tilde{V}}{s}.  \label{eleqom1}
\end{equation} 

Evidently, a separation of the CM from the relative motion is provided if and only if $\pd{\tilde{V}}{S}$ is exclusively a function of $S$  which is equivalent to
\begin{equation}
 \frac{\partial^2 \tilde{V}}{\partial s \partial S}=0 \Leftrightarrow \tilde{V}(S,s)=V_1(S)+V_2(s)  \label{potsep2}
\end{equation}
with $V_1, V_2$ being arbitrary functions of $S$ and $s$ respectively.

In order to analyze this condition further, we  technically
have to distinguish the two cases of a regular and a singular potential $\tilde{V}(S,s)$ at $s=0$.\newline
\\
 For a regular potential, we obtain from (\ref{potsep1})
that $\tilde{V}(S,0)=V(0)$. Condition (\ref{potsep2}) then yields
\[V_1(S)=V(0)-V_2(0)=const. \Rightarrow \pd{V_1}{S}=\pd{\tilde{V}}{S}=0.\] \newline
For a potential with a singularity at $s=0$, as in the typical case of  Coulomb interaction,
 choosing an infinitesimal value $\varepsilon>0$ for the $s$ coordinate, we are led through (\ref{potsep1}) to
\begin{eqnarray}
 \tilde{V}(S,\varepsilon) &=& V\left(\abs{\vec{r}(S+\frac{m_2}{M}\varepsilon)-\vec{r}(S-\frac{m_1}{M}\varepsilon)}\right)\nonumber \\
&=& V\left(\abs{\vec{r}(S)+\frac{m_2}{M}\varepsilon \vec{t}(S)-\vec{r}(S)+\frac{m_1}{M}
\varepsilon \vec{t}(S)}\right)\nonumber \\
&=& V\left(\abs{\varepsilon \vec{t}(S) }\right) = V\left(\varepsilon\right), \label{potsep3}
\end{eqnarray}
since $\abs{\vec{t}(S)}=1$.
Thus, for arbitrary $S,\tilde{S}$
\[\tilde{V}(S,\varepsilon)=\tilde{V}(\tilde{S},\varepsilon)=V(\varepsilon) \]
and Eq. (\ref{potsep2}) for $s=\varepsilon$ leads to
\[V_1(S)=V_1(\tilde{S})=V(\varepsilon)-V_2(\varepsilon) ~\forall S,\tilde{S}~ \Rightarrow \pd{V_1}{S}=\pd{\tilde{V}}{S}=0.\]

Therefore both cases lead to the condition $\pd{\tilde{V}}{S}=0$ and  we conclude that the CM and relative 
motion mutually separate for a potential that depends only on the
inter-particle Euclidean distance  (Eq. (\ref{potsep1})) if and only if $\pd{\tilde{V}}{S}=0$.
Furthermore, this is a necessary and sufficient condition for the conservation of the total momentum 
 \[P=\pd{L}{\dot{S}}=M\dot{S}\] as 
follows from (\ref{fueq3}) and the EL equations (\ref{eleqom1}), yielding a free particle motion for the CM.

Introducing the function $R(s_1,s_2)=\abs{\vec{r}(s_1)-\vec{r}(s_2)}$, with $s_1=S+\frac{m_2 }{M}s$, $s_2=S-\frac{m_1}{M}s$ and $V'(R)=\d{V}{R}\big|_{R=R(s_1,s_2)}$ we obtain
\begin{eqnarray}
 \pd{\tilde{V}}{S}&=& V'(R) \pd{R}{S} \nonumber \\
&=&V'(R) (\partial_{s_1} R(s_1,s_2)+\partial_{s_2} R(s_1,s_2))  \nonumber \\
\end{eqnarray}
 and  we are thus led to the conclusion that a conservation of the total momentum as well as a separation of the CM from the relative coordinate is provided for interacting particles if and only if
\begin{equation}
 \pd{R}{S}=\partial_{s_1} R(s_1,s_2)+\partial_{s_2} R(s_1,s_2)=0, ~\forall s_1,s_2 \in \mathbb{R}. \label{con0}
\end{equation}

The results of \cite{Schmelcher11} indicate that for the confining manifold being a homogeneous helix, i.e. a helix with a constant radius and pitch,
the CM motion is separated from the relative one. In fact, the homogeneous helix - including also the limiting cases of the straight line 
and the circle - is the only curve allowing for such a separation,
 as follows from the proposition below.
 \newline
\\
\textbf{Proposition:} \textit{Condition (\ref{con0}) holds for a smooth, regular curve $\vec{r}(s)$ that is either closed or extends to infinity 
and is parametrized by its arc length $s \in \mathbb{R}$
if and only if  the curve is a homogeneous helix. }
 \newline
\\
The proof of this proposition is provided in the Appendix.

\begin{center}
 { \bf{III. TWO CHARGED PARTICLES IN AN INHOMOGENEOUS HELICAL TRAP}} 
\end{center}
In  the following  we study the classical dynamics of two identical charged particles confined in a modified helix.
The modification consists of a hump, i.e.  a local change of Gaussian form in the radius  (Eq. (\ref{hum1})). The interaction between the particles is given by a repulsive Coulomb potential
\[V(\abs{\vec{r}_1-\vec{r}_2})=\frac{\lambda}{\abs{\vec{r}_1-\vec{r}_2}},\]
with $\lambda>0$.
We explore in particular the effects due to the coupling of the CM and relative motion in the presence 
of the helical hump.

\begin{figure}[htbp]
\begin{center}
\includegraphics[width=8.6cm]{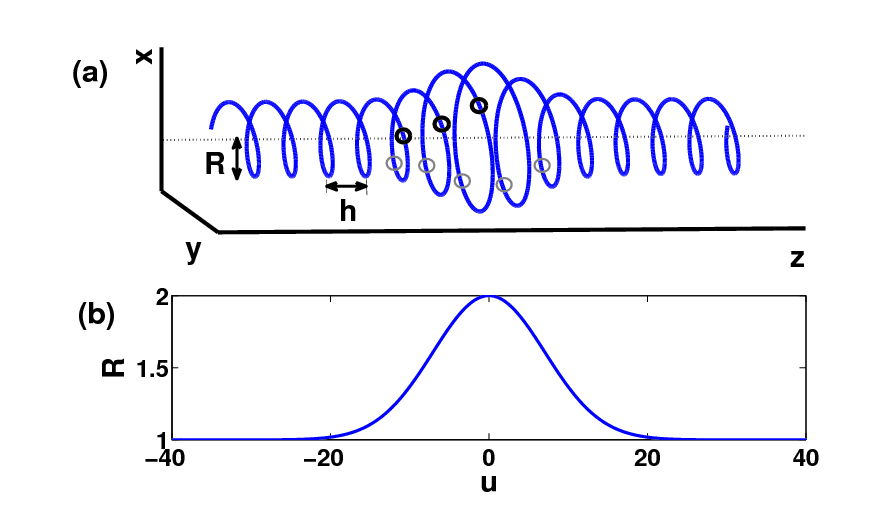}
\end{center}
\caption{\label{fhel1} (color online). (a) Helix with a local modification of  the radius $R(u)$ and a pitch $h$, as given by Eq. (\ref{hel1}).
 (b) Local modulation of the radius  as a function of the parameter $u$.}
\end{figure}

The inhomogeneous helix  parametrized by the angle parameter $u$ is given by:
\begin{equation}
\vec{r}(u)=(R(u) \cos (u), R(u) \sin (u), \frac{h}{2 \pi} u) \label{hel1}
\end{equation}
with 
\begin{equation}
R(u)=1+\epsilon \exp[-cu^2],\label{hum1}
\end{equation}
where both the modified radius $R(u)$ and the pitch of the helix $h$ have been scaled with the radius of the corresponding uniform helix $R_0$.
We use for the inhomogeneous helix the parameter values $\epsilon=1,~c=0.01,~ h=0.4 \pi$. Figure \ref{fhel1}  depicts the shape of such a helix and the localized 
radial modulation.

Since the particles are identical, $m_i=m$, we can choose dimensionless units by rescaling all quantities with $m$, $\lambda$ and $R_0$, i.e. introducing $\tilde{m}=\tilde{\lambda}=1$ and
\[\tilde{x}=\frac{x}{R_0}, ~\tilde{t}=t\sqrt{\frac{\lambda}{mR_0^3}},~ \tilde{H}=\frac{H R_0}{\lambda},~\tilde{p}=p\sqrt{\frac{R_0}{m \lambda}}\]
In the following we omit for simplicity the tilde.

 Performing then a Legendre 
transformation with $p_i=\pd{L}{\dot{u}_i}$ we obtain from (\ref{lanui}) the Hamiltonian
\begin{equation}
H(\{u_i, p_i\})=\frac{1}{2}\sum_{i=1}^2 \frac{p_i^2}{(\partial_{u_i}\vec{r}(u_i))^2}+\frac{1}{\abs{\vec{r}(u_1)-\vec{r}(u_2)}}\label{hmlt1}
\end{equation}
From this we deduce the equations of motion $\dot{u}_i=\pd{H}{p_i}, \dot{p}_i=-\pd{H}{u_i}$, which we solve numerically for different initial  conditions
with a Runge-Kutta method of fourth-fifth order with a variable time step size (ODE45). 
In order to study the dynamics in terms of CM and relative motion,
it is desirable to have a Hamiltonian with a kinetic energy term of Cartesian form. This is  achieved under the arc length parametrization  (\ref{alp1})
and leads to
\begin{equation}
H(\{s_i, \dot{s}_i\})=\frac{1}{2}\sum_{i=1}^2 \dot{s}_i^2+\frac{1}{\abs{\vec{r}(s_1)-\vec{r}(s_2)}}. \nonumber
\end{equation}
The $s_i$ are obtained (numerically) via Eq. (\ref{alp1}).
Introducing CM $S=\frac{s_1+s_2}{2}$ and relative coordinates $s=s_1-s_2$
yields the Hamiltonian
\begin{equation}
H(S,s, \dot{S},\dot{s})=\dot{S}^2+\frac{\dot{s}^2}{4} +\frac{1}{\abs{\vec{r}(u_1(S,s))-\vec{r}(u_2(S,s))}}\label{hmlt2}
\end{equation}
and the corresponding equations of motion:
\begin{eqnarray}
\ddot{S}&=&-\frac{1}{2}  \frac{\partial}{\partial S} \frac{1}{\abs{\vec{r}(u_1(S,s))-\vec{r}(u_2(S,s))}}  \nonumber \\
 \ddot{s}&=&-2  \frac{\partial}{\partial s} \frac{1}{\abs{\vec{r}(u_1(S,s))-\vec{r}(u_2(S,s))}}. 
\label{hmlteqom1}
\end{eqnarray}
We clearly observe here the coupling between $S$ and $s$ in the potential term. 
In the case of the uniform helix \cite{Schmelcher11} the arc length integral can be solved analytically  and the Hamiltonian can be written explicitly as:
\begin{equation}
H(s, \dot{S},\dot{s})=\dot{S}^2+\frac{\dot{s}^2}{4} +\frac{1}{\sqrt{2\left(1-\cos (\frac{s}{a})\right)+(\frac{h}{2 \pi a})^2 s^2}},\label{uhmlt}
\end{equation}
with $a=\sqrt{1+\left(\frac{h}{2\pi}\right)^2}$.

For understanding the dynamics it is crucial to analyze the properties of the potential $V(S,s)$.
Obviously, we have $ \lim_{R \rightarrow \infty} V(R)=0$.
We focus first on the uniform helix for which $V=V(s)$ (see Eq. (\ref{uhmlt})) 
 and thereafter we consider the case of the coupling of the CM and relative motion. 
Figure \ref{upot} below shows the behaviour of this potential curve for $s<20$.
 We identify three potential wells which can support bound states and become shallower as $s$ increases.

\begin{figure}[htbp]
\begin{center}
\includegraphics[width=8.6cm]{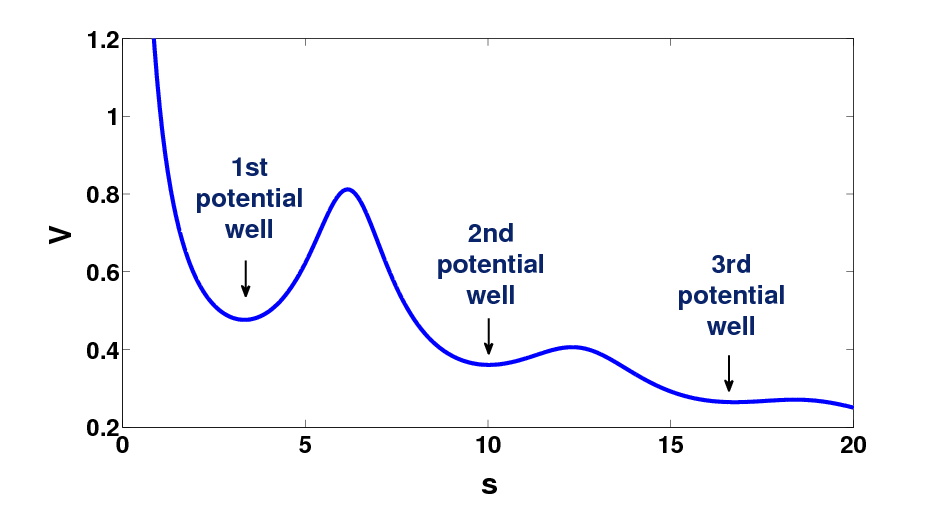}
\end{center}
\caption{\label{upot} (color online). Potential curve for the uniform helix with parameters $h=0.4 \pi$ and $R=1$. We observe three potential wells located at $s=3.34, 10.00$ and $16.75$
with minimum values $V=0.48, 0.36$ and $0.26$ respectively. }
\end{figure}

\begin{figure}[htbp]
\begin{center}
\includegraphics[width=8.6cm]{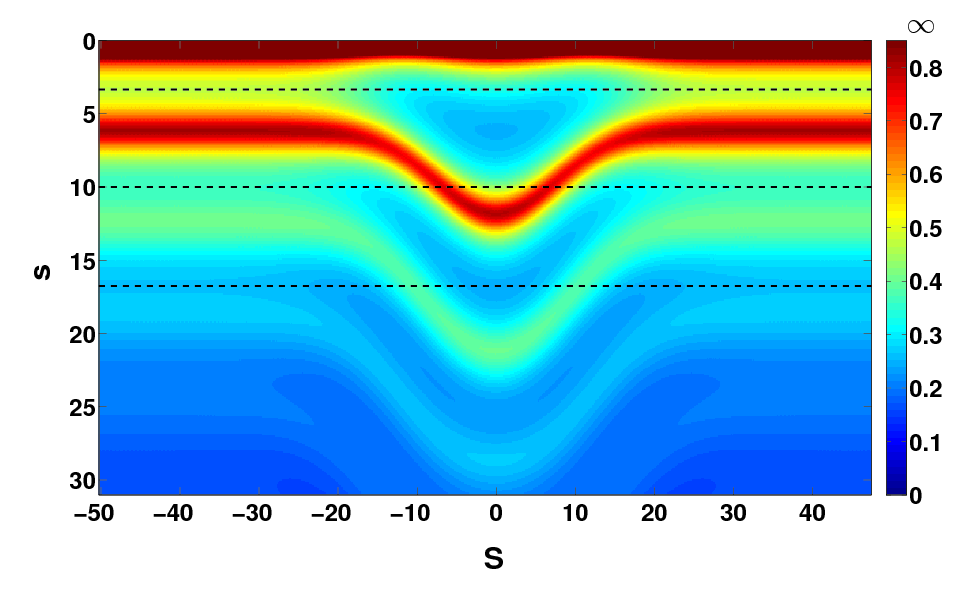}
\end{center}
\caption{\label{hpot} (color online). Contour plot of the potential $V(S,s)$ for the inhomogeneous helix. The dashed lines represent the positions of the minima of the three wells
for the homogeneous case. The effect of the local modification of the radius is evident 
for $\abs{S}\lesssim 30$. }
\end{figure}

The potential $V(S,s)$, taking into account the hump, is illustrated in Fig. \ref{hpot}. Since  it depends on both the CM $S$ and the relative $s$ coordinate, it  
represents a two-dimensional potential landscape.

We clearly observe two regions with a distinct behaviour. The first, for large values of the CM coordinate $\abs{S}\gtrsim 30$, presents a uniform behaviour, approximately
 independent of $S$.
It is affected only by the relative coordinate $s$ in the same way as the potential of the homogeneous helix (Fig. \ref{upot}), thereby presenting three wells for $\abs{s} \approx 3,10,17$.
 In this uniform
 domain the CM and the relative motion are thus decoupled.
The second region, for $\abs{S} \lesssim 30$, presents a 
strong dependence on the CM coordinate and  thus constitutes a regime of strong coupling. 
The reader should note that the arc length (\ref{alp1}) is taken w.r.t. the centre of the hump and consequently regions with small $S,s$ correspond to small $s_i$ and  lie in the inhomogeneous region of the helix. 

Two effects are evident: each potential well becomes deeper and
 the contour lines bend in the regime of the inhomogeneity. Concerning the potential barriers, their maximum value decreases by $\sim 8 \%$ at the sides of the inhomogeneous region,
 whereas for $S\approx 0$ it retains the value of the homogeneous regime. All these effects can be  explained by the modulation of the radius of the helix as
discussed below.  

\begin{figure}[htbp]
\begin{center}
\includegraphics[width=8.6cm]{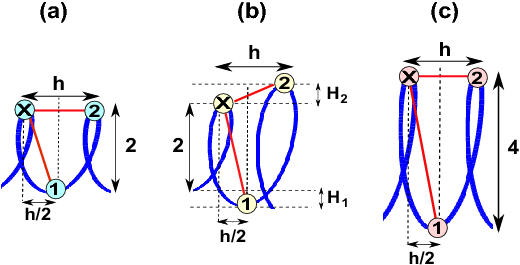}
\end{center}
\caption{\label{hepot} (color online). Euclidean distances of the particles for the minimum  $(\circx, \circone)$ and the maximum $(\circx, \circtwo)$ potential configurations of the
 first potential well for three different regions of the inhomogeneous helix:
(a) the uniform domain of the helix with pitch $h$ and radius $R=1$, (b) the left side of the inhomogeneous region where the radius increases by $H_1$ at the bottom and by
$H_2$ at the top, whereas the pitch remains the same, (c) the central part of the  inhomogeneous domain $S\approx 0$, which since the modulation of the radius is stationary
can be approximately treated as a part of a uniform helix with the same pitch $h$, but double radius $R=2$. }
\end{figure}

\textit{Modulation of the radius and potential landscape.} Since the pitch of the helix is much smaller than its circumference $(h<2 \pi R)$ in both the homogeneous and the inhomogeneous regime, 
the maximum and the minimum potential configurations  occur 
for approximately constant values of the relative angle parameter $\tilde{u}=u_1-u_2$,  
namely for $\tilde{u}_{max}=2k\pi$ and $\tilde{u}_{min}=(2k-1)\pi, ~k \in \mathbb{Z}$ \cite{Schmelcher11}, 
which for the first well ($k=1$)
correspond respectively to particles separated by one or half a winding of the helix (Fig. \ref{hepot}). The Euclidean distance between the particles at the  minimum configuration increases
substantially with the increment of the radius  (Fig. \ref{hepot} (b), (c)), reaching its maximum value at $S \approx 0$ (Fig. \ref{hepot} (c)) thereby resulting in a strong increase of the 
potential depth.

For the maximum configurations, the Euclidean distance increases as well off the centre of the inhomogeneous region but, as it is shown in Fig \ref{hepot} (b), 
this increment is small compared to that of the minimum configuration. At $S\approx 0$, the rate of change of the radius becomes small enough for the radius
to be considered constant with twice the value of its   
homogeneous asymptotics (Fig. \ref{hepot} (c)). However the pitch $h$ remains the same, resulting
in the same Euclidean distance between the particles of the maximum configuration and thus leading to the same maximum potential values as that of the uniform domain (Fig. \ref{hepot} (c)).
The generalization to other potential wells (second, third) is evident in the regime $S \approx 0$: for the case of the maximum configuration both particles are shifted by the same distance in the same direction whereas
 for the case of the minimum configuration by the same distance in opposite directions. This fact results both in an unaltered maximum value of the potential barrier and in a
considerable increase of the potential depth.
 
 Finally we note that  the contour lines of the potential bend inside the inhomogeneity towards larger $s$ values as compared to the
uniform domain. This effect is more pronounced for larger
relative distances, i.e for the third well as shown clearly in Fig. \ref{hpot}.

In the following sections we will not discuss the dynamics of the system in terms of the trajectories $t \mapsto s_i (t)$ of two individual 
particles but rather in terms of that of a fictitious particle 
with two degrees of freedom - $S$ and $s$ - moving in the 2D potential of  Fig. \ref{hpot}. 
This interpretation is suggested by the form of the Hamiltonian (\ref{hmlt2}), which provides us with the respective 
equations of motion (\ref{hmlteqom1}). Note however, that these two degrees of freedom have different effective masses, a fact that needs to be taken into account when investigating the dynamics
 of the fictitious particle in terms  of the potential gradients in the CM and the relative direction.

\begin{center}
 { \bf{IV. SCATTERING OFF THE HELICAL HUMP}} 
\end{center}

We analyze now the scattering behaviour of a bound pair of charged particles confined in the inhomogeneous helix that has been described above. 
We assume that the particles start in the uniform domain, i.e. for $S \ll -30$ with $\dot{S} > 0$. We introduce $S_h$ as the value of $\abs{S}$ after which the
helix as well as the potential are considered uniform. Specifically we choose $S_{h}\approx 35.6$  for which the radius is identical to that of the homogeneous helix within $0.1 \%$.
 The particles are further assumed to be initially bound, so their relative coordinate $s$ lies within the region of one of the three wells   
discussed in Sec. IV. As they pass through the inhomogeneous region, energy is transfered  between the CM and the relative degree of freedom due to the coupling. This transfer 
can lead to dissociation of the particles, which is reflected in the very low values of the interaction potential ($V \rightarrow 0$) at the end of the propagation (i.e. for $S \gg 30$).

We will initially discuss the case where the particles start with zero relative velocity $\dot{s}=0$ at the minimum of    
each of the three wells (Subsec. A) and then examine further the case of the first potential well for different  initial conditions (Subsec. B).

\begin{center}
 { \bf{A. Initial conditions with zero relative momentum}} 
\end{center}
 \textit{First potential well.} The particles are placed in the homogeneous domain ($S \ll 0$) of the helix, in  the minimum of the first potential well $s=3.34$ (Fig. \ref{upot}), with $\dot{s}=0$.
We vary the initial values of the CM kinetic energy $T_S=\dot{S}^2$,  over several orders of magnitude. 
The different initial conditions are propagated for a time period
 $t=600$.
After that time the particles have passed the region of inhomogeneity of the helix, the scattering process is in its asymptotic regime, and we can record the final values of the potential energy $V$, the relative energy $E_s=\frac{\dot{s}^2}{4}+V$, 
and the relative coordinate $s$. If the final value of $V$ lies within the first potential well, i.e. $0.48<V<0.81$, then the particles have remained bound, whereas if 
$V$ approaches zero they  have dissociated through the scattering, which is also ensured by large values of the relative coordinate.

\begin{figure}[htbp]
\begin{center}
\includegraphics[width=8.6cm]{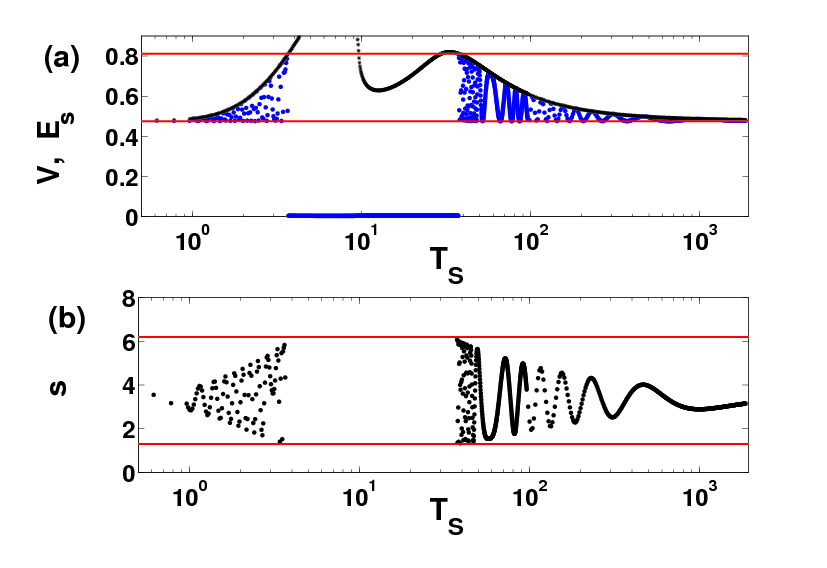}
\end{center}
\caption{\label{1stwe} (color online) Overview of final bound and unbound states for initially bound states started in the first potential well:
 (a) Final potential values $V$ (blue dots) and final relative energy values $E_s$ (black dots) for different initial CM kinetic energies $T_S$.
The red lines represent the boundary potential values (minimum of well, maximum of barrier) of the first potential well.
 (b) Final relative coordinate values $s$ (black dots) for different initial CM kinetic energies $T_S$.
The red lines represent the boundary  values of those $s$ which lie within the first potential well.}
\end{figure}

We clearly observe  in Fig. \ref{1stwe} (a) two regimes of finally bound  configurations: $0 \leq T_S \leq 3.83$ and $T_S \geq 38.86$,  separated by a region of dissociation $3.83 \leq T_S \leq 38.86$.
For small $T_S$, below a critical value $T_{S_{c1}}$ it is expected that an energy transfer between the CM and the relative degree of freedom would not provide sufficient energy 
so that the particle can overcome the potential barrier.  After   $T_{S_{c1}} \approx 3.83$ dissociation becomes possible and it indeed occurs. However, the dissociation regime stops at a second 
critical value
$T_{S_{c2}} \approx 38.86$, a fact that although counterintuitive from the point of view of the possible energy supply, can be explained  by the limited range of the inhomogeneous region
 ($\abs{S} \leq 30)$.
For very high CM velocities $\dot{S}$, the particles get through the inhomogeneity very fast, allowing for a very short interaction time only. The effect of the coupling is 
therefore very restricted,
 prohibiting a substantial energy transfer. In other words, the particles' motion is almost unaffected by the presence of the hump due to their large velocities. 
In the regime of bound states the change of $T_S$ induces a change of the $s$-oscillation phase at  
the end of the propagation ($t=600$) leading to an oscillatory pattern of the final values of $V$ and $s$. Another interesting feature of
 Fig. \ref{1stwe} (a), is the behaviour of the  final relative energy $E_s$. In the middle of the dissociation region, it acquires values less than that of the potential barrier 
$V_{max}\approx 0.81$, a fact that will be analyzed in Subsec. B.

\begin{figure}[htbp]
\begin{center}
\includegraphics[width=8.6cm]{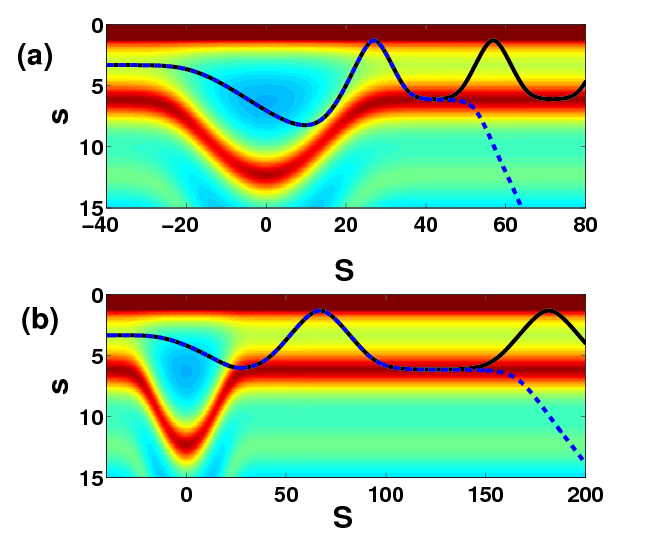}
\end{center}
\caption{\label{1stwtr} (color online) Finally bound (solid black line) and dissociated (dashed blue line) trajectories near:
 (a) the first transition point $T_{S_{c1}} \approx 3.83$,
 (b) the second transition point $T_{S_{c2}} \approx 38.86$. The trajectories in each case ((a),(b)) differ in their CM velocities only by $0.001$.}
\end{figure}

Let us next explore the behaviour of the trajectories for $T_S$ close to the critical values  $T_{S_{c1}}, T_{S_{c2}}$ which will be referred to in the following as  transition points.
Our results are presented in  Fig. \ref{1stwtr}.  In both cases (Figs. \ref{1stwtr} (a),(b)) a sharp transition from a bound to a dissociated final state occurs when $T_S$ is fine tuned. This is depicted
in the form of the corresponding trajectories which are essentially on top of each other for $S$ less than a critical value $S_c$. This value is 
much larger for the second transition point with $T_{S_{c2}}> T_{S_{c1}}$, a fact that can be attributed to the larger value of the CM velocity. There is 
an evident transfer of energy to the relative degree of freedom depicted in the very large amplitude of the $s$-oscillation of the fictitious particle for bound trajectories,
after the scattering. With a slight increment of this transfer  the states dissociate after an oscillation. The trajectory of the fictitious particle is deflected inside the 
hump following the curved topology of the potential landscape until it comes across a large value of the potential barrier where it becomes reflected backwards. From then on,
 it continues its 
regular path in the right homogeneous domain without any further energy transfer. As expected, the motion of the particle is much less affected (smaller angle of deflection)
 by the presence of the inhomogeneity for larger CM velocity (Fig. \ref{1stwtr} (b)) due to its inertia.

\begin{figure}[htbp]
\begin{center}
\includegraphics[width=8.6cm]{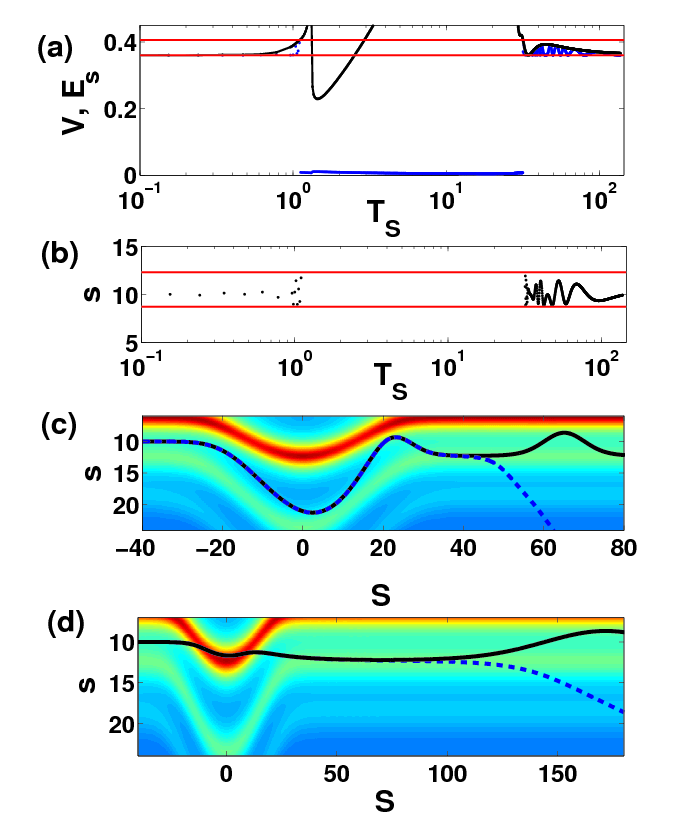}
\end{center}
\caption{\label{2we} (color online) 
 (a)-(b)  Same as in Fig. \ref{1stwe}, but for particles starting in the second potential well.
(c)-(d) Same as in Fig. \ref{1stwtr} but with: (c) $T_{S_{c1}} \approx 1.15$, (d) $T_{S_{c2}} \approx 33.54$.}
\end{figure}

 \textit{Second potential well.} We now place the particles in the minimum of the second potential well $s=10.00$ (Fig. \ref{upot}) at the left homogeneous region, with zero 
relative velocity. Varying the CM kinetic energy $T_S$ we observe again (Figs. \ref{2we} (a),(b))  a region of dissociation $1.15 \leq T_S \leq 33.53$ surrounded by regions
 of  bound states, yielding two transition points:  $T_{S_{c1}}=1.15$ (Fig. \ref{2we} (c)) and $T_{S_{c2}}=33.53$ (Fig. \ref{2we} (d)).
 The dissociation region is overall shifted to lower values of $T_S$, compared to our previous results for the first potential well.
 From an energetical point of view, the shift of the first transition point $T_{S_{c1}}$ is expected since the potential barrier
lowers, allowing for dissociation with less energy transfer. However, this argument alone would lead to a shift of $T_{S_{c2}}$ to larger values, contrary to what is observed here. 
The suppression of $T_{S_{c2}}$ seems to be a result of the bending of the potential landscape inside the hump. In particular,
the straight line indicating the minimum of the second potential well in the homogeneous regime, passes through the first potential well 
close to its barrier (Fig. \ref{hpot}). Trajectories with high enough CM kinetic energy $T_S \geq T_{S_{c2}}$, encounter this barrier and 
are forced to crest it (Fig. \ref{2we} (d)), a fact
that reduces abruptly the amount of the energy transfer and leads to extended binding. This effect is more pronounced in the case of the third potential well as discussed below. 



\begin{figure}[htbp]
\begin{center}
\includegraphics[width=8.6cm]{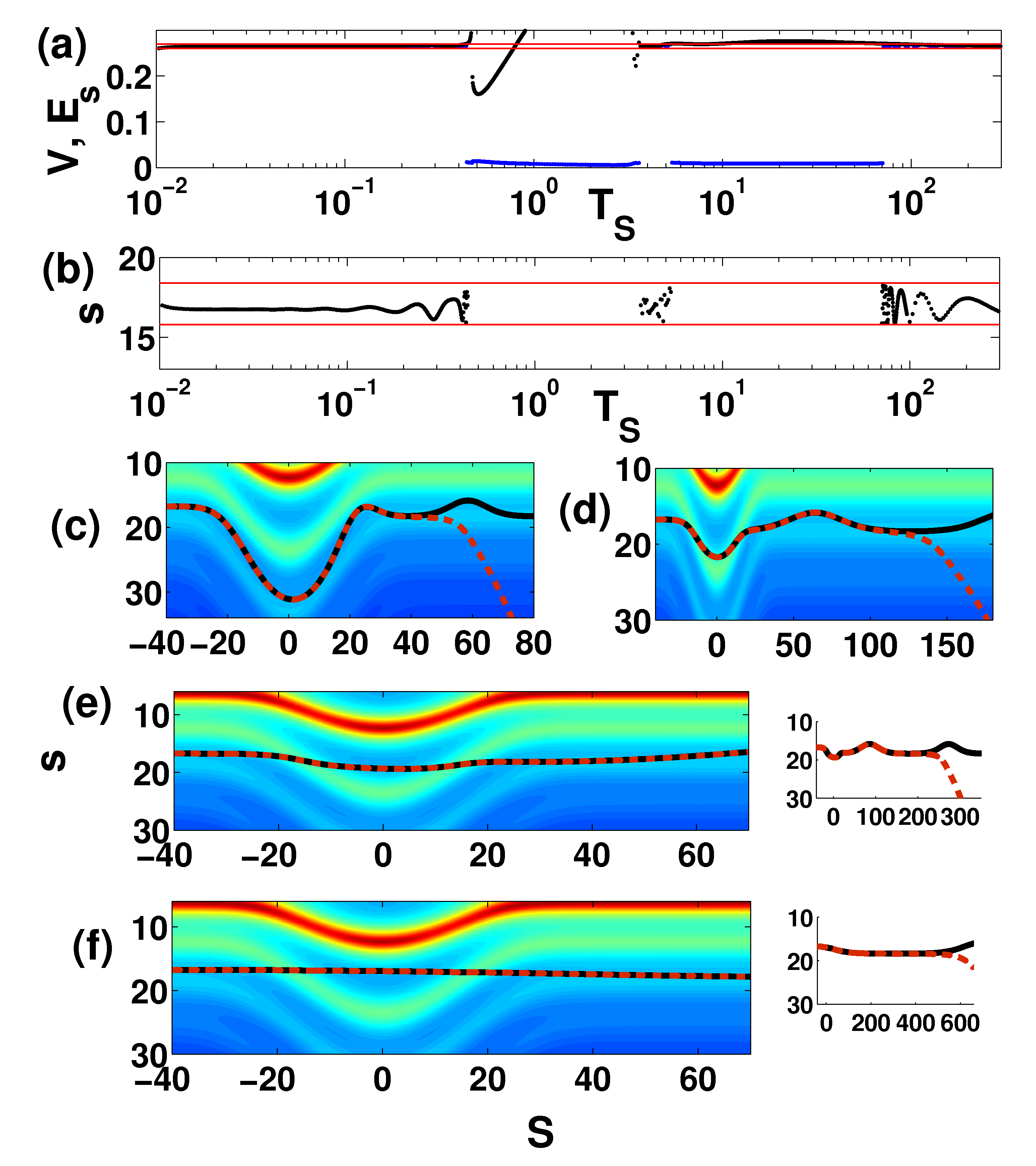}
\end{center}
\caption{\label{3pwe1} (color online)  (a)-(b)  Same as in Fig. \ref{1stwe}, but for particles starting in the third potential well.
(c)-(f)  Same as in Fig. \ref{1stwtr} but with:
 (c) $T_{S_{c1}} \approx 0.46$,
 (d) $T_{S_{c2}} \approx 3.76$,
(e)  $T_{S_{c3}} \approx 5.59$,
(f)  $T_{S_{c4}} \approx 70.35$.
The small subfigures in (e) and (f) present the respective trajectories for large values of $S$, following them up to the point where the dissociative and bound trajectories separate
from each other.}
\end{figure}

\textit{Third potential well.} Similarly to the previous cases, we now place the particles in the minimum of the third potential well $s=16.75$ (Fig. \ref{upot}) at the left homogeneous region, 
again with $\dot{s}=0$. A variation of the CM kinetic energy $T_S$ (Figs. \ref{3pwe1} (a), (b)),  provides us surprisingly with two distinct dissociation regimes
 $0.46 \leq T_S \leq 3.76,~5.59 \leq \dot{S} \leq 70.35$, separated by a small region of bound states  ($3.76 \leq T_S \leq 5.59$), leading to
four transition points $T_{S_{c1}}=0.46,~T_{S_{c2}}=3.76,~T_{S_{c3}}=5.59,~T_{S_{c3}}=70.35$ (Figs. \ref{3pwe1}  (c)-(f)).

This fact is  a direct result of the bending of the potential landscape, which affects mainly the larger relative coordinates $s$, i.e. the third well. In particular,
as depicted in Fig. \ref{hpot} the straight line of the minimum of the third potential well, passes, inside the hump, above the minimum  of the second potential well.
Due to the shallowness of the third potential well, only a very small amount of energy transfer $\sim 0.01$ is needed for the particles to overcome the barrier and dissociate,
a fact that shifts the first transition point $T_{S_{c1}}$ to low values. 
When the fictitious particle has enough CM kinetic energy ($T_s \geq T_{S_{c2}}$), it crests the barrier of the second well, but since it gets directly
 deflected within it,
it cannot reach its inner region and the minimum (Fig. \ref{3pwe1} (d)). This is similar to the case of the second transition point of the second potential well (Fig. \ref{2we} (d))
and as in there, it is followed by a regime of bound states. However, if the CM velocity gets large enough ($T_{S_{c3}}=5.59$), the fictitious particle is less deflected
and can reach the region of the minimum of the second potential well (Fig. \ref{3pwe1} (e)), allowing  for further energy redistribution between the two degrees of freedom. 
Thus, a second dissociation region occurs, which extends to very high values of $T_S \approx 70$, a fact that can also be attributed to the very small height of the potential barrier.
Nevertheless, even this height  cannot be overcome, when the fictitious particle acquires
CM kinetic energy larger than $T_{S_{c4}}$, since the dwell time becomes very small, leading again to bound trajectories as in the cases of the other wells.

 For this potential well,  sharp transitions from a bound to  an unbound state occur too at the four transition points as shown in Fig. \ref{3pwe1}(c)-(f). 
It is evident that at the fourth transition point with a large value of $T_S$, the motion of the fictitious particle is only slightly affected by the inhomogeneity, 
tending to a straight line (Fig. \ref{3pwe1} (f)). 

We emphasize that since the equations of motion of the system (Eq. (\ref{hmlteqom1})) possess a time reversal symmetry, the transitions 
from bound to unbound states can be directly mapped into transitions from free states to bound ones. The creation of bonds through scattering is surprising, especially in view of the 
fact that the particles interact via a repulsive Coulomb potential. For these reasons we find it interesting to examine this process further below.

\begin{center}
 { \bf{B. Initial conditions with nonzero relative velocity in the first potential well.}} 
\end{center}

\begin{figure*}[htbp]
\begin{center}
\includegraphics[width=17cm]{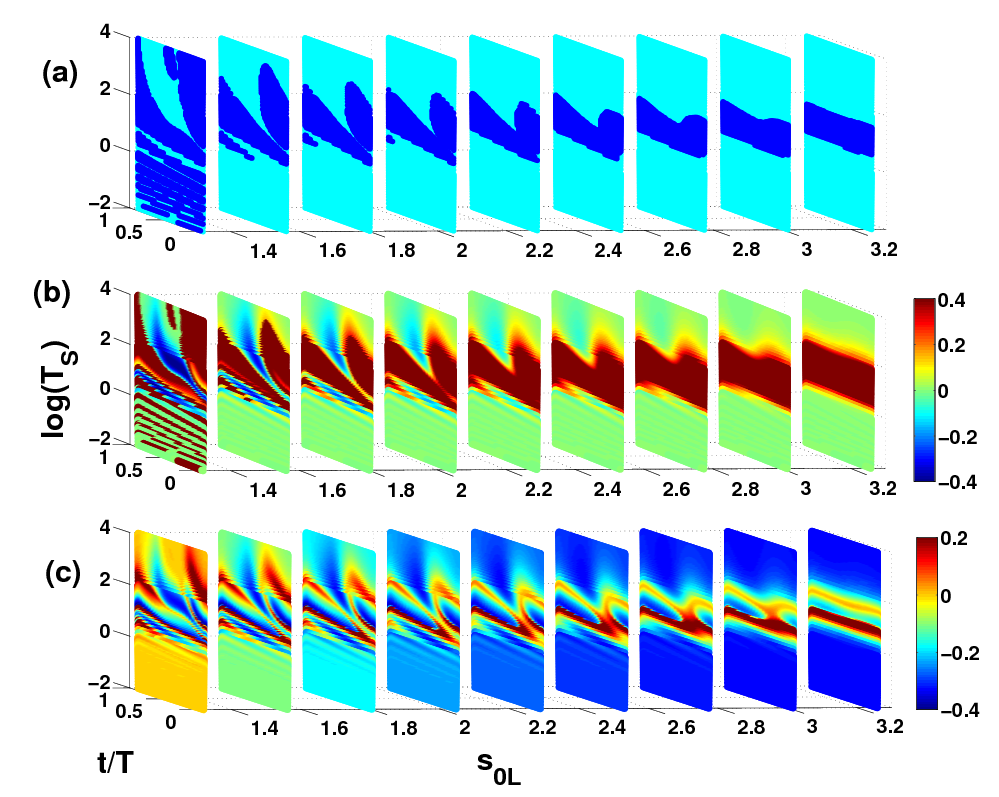}
\end{center}
\caption{\label{scat1we} (color online) (a) States that after scattering remain bound (cyan regions) and states that are led to dissociation (blue regions), 
(b) color encoded values of the relative  difference between the final and the initial maximum kinetic energy of  the relative motion $(T^s_{max_f}-T^s_{max_i})/E_s$,
(c) color encoded values of the difference between the final relative energy  $E_{s_f}$ and the maximum value of the potential barrier of the first well
$V_{max}$ for different $s_{0L}$ (X axis), $\frac{t}{T}$ (Y axis) and $\log (T_S)$ (Z axis). Note that in case (b) the zero value is within the green region whereas in case (c) 
within the orange one.
}
\end{figure*}

We  investigate now  the scattering  for arbitrary initial conditions, focusing on the first potential well.
When the particles are in the uniform domain of the helix, inside the first well with a nonzero relative velocity, the fictitious particle
performs an oscillation in the relative coordinate $s$. The phase of this oscillation when the fictitious particle enters the hump affects the energy transfer between the CM and the 
relative motion. Moreover, the value of the relative initial energy $E_s$ plays a crucial role in determining which states become dissociated, since states with higher
$E_s$  require less
 amount of energy transfer in order to overcome the potential barrier. Thus, for a complete description of the scattering process, we need except from the initial center
 of mass kinetic energy $T_S$ to specify two other parameters,
namely the initial relative energy and the phase of the relative oscillation.

For reasons of convenience we assume that the fictitious particle starts at a point $1.3<s_{0L} \leq s_{min}=3.34$  of the first potential well in the uniform domain 
with zero relative velocity $\dot{s}=0$. In
other words $s_{0L}$ is the left turning point of the oscillation in the relative coordinate and is related to the total relative energy by $E_s=V(s_{0L})$, with $V$ being the potential 
of the homogeneous regime given by  Eq. (\ref{uhmlt}). We denote the right turning point for the same energy with $s_{0R}$.

We represent  the phase of the oscillation, by the parameter $0 \leq \frac{t}{T} <1$
which stands for the fraction of the period of the relative oscillation
 \[T=2 \int_{s_{0L}}^{s_{0R}} \frac{ds'}{\sqrt{2 (E_s-V(s')})}\]
at which the particles enter the hump. In such a way $\frac{t}{T}=0$ corresponds to particles at $s_{0L}$ with $\dot{s}=0$ at the entrance point $S_{hL}=-S_h$,
whereas $\frac{t}{T}=0.5$ corresponds to particles at $s_{0R}$ with $\dot{s}=0$. 
This parameter can be adjusted by changing the initial CM coordinate $S$ in the homogeneous region, while keeping $\dot{S}$ fixed.  Due to our genuine interest in the scattering properties with varying phase, the absolute
phase dependence induced by the arbitrariness of $S_h$ is rendered  irrelevant.

Our results are presented in Fig. \ref{scat1we} for nine representative values of $s_{0L}$, ranging from energies close to the potential minimum ($s_{0L} \approx 3.3$) to close
 to the potential barrier  height  ($s_{0L} \approx 1.3$). Each such value produces a slice which imprints the dependence of the property under consideration on the other two parameters:
$T_S$ and $\frac{t}{T}$. 

Figure  \ref{scat1we} (a) provides us with the finally bound and unbound states for the different initial conditions. For $s_{0L} \approx 3.3$, close to the minimum,
we observe that  the phase of oscillation $\frac{t}{T}$ does not  affect the behaviour of the system, as expected, and we regain the results of Subsec. A with a single dissociation region
of a rectangular shape for different CM kinetic energies  $T_S$. The shape of this regime is deformed as we go to higher relative energies (smaller $s_{0L}$) and  it develops a dip.
 By increasing further $E_s$  the dissociation area breaks into two parts  
for a certain regime of initial phases $\frac{t}{T}$, providing us with two dissociation regions with varying $T_S$. 
As we approach the threshold energy for passing the potential barrier ($s_{0L} \approx 1.3$)  we observe an alternating sequence of bound and
dissociation regions, even at very low CM kinetic energies. This  is  a surprising feature  which makes the dissociation process sensitive to even small changes of the underlying parameters, such as the initial value of $T_S$ in the
scattering process.

In Fig.  \ref{scat1we} (b) we present our results for the relative amount of gain or loss of the  maximum relative kinetic energy $(T^s_{max_f}-T^s_{max_i})/E_s$ through the scattering.
Clearly the regions of high positive  $(T^s_{max_f}-T^s_{max_i})/E_s$ match exactly with the dissociation regions of Fig. \ref{scat1we} (a). Most regimes show almost zero total gain of kinetic energy,
 but surprisingly enough there are also regimes where the kinetic energy of the relative motion is decreased after the scattering. These regions of loss predominately appear  between 
the different regimes of dissociation and are characterized by particles becoming more tightly bound in the course of scattering. A further interesting observation can be made in  Fig.  \ref{scat1we} (c) which shows 
 the difference $E_{s_f}-V_{max}$ of the final relative energy  $E_{s_f}$ and the maximum value of the potential barrier of the first well
$V_{max}$. In particular, this difference is negative not only for the finally bound states, but also for some of the finally dissociated ones. The dissociation regions consist of states
 with  $E_{s_f} \geq V_{max}$ at their boundaries and of ones  with  $E_{s_f} \leq V_{max}$ at their center. 
The dissociated states with $E_{s_f} \leq V_{max}$ might seem counterintuitive but
as it will be shown below, these result from trajectories for which the particles dissociate within the hump, where the coupling of the CM and relative motion is still
substantial.
 Since the potential  barrier in this region is bent (Fig. \ref{hpot}) the fictitious particle 
overcomes it with its total amount of energy $E=T_s+T_S+V(S,s)$, and thus its final relative energy can be less than $V_{max}$.
 Such a phenomenon has already been encountered in our investigations of Subsec. A.

\begin{figure}[htbp]
\begin{center}
\includegraphics[width=8.6cm]{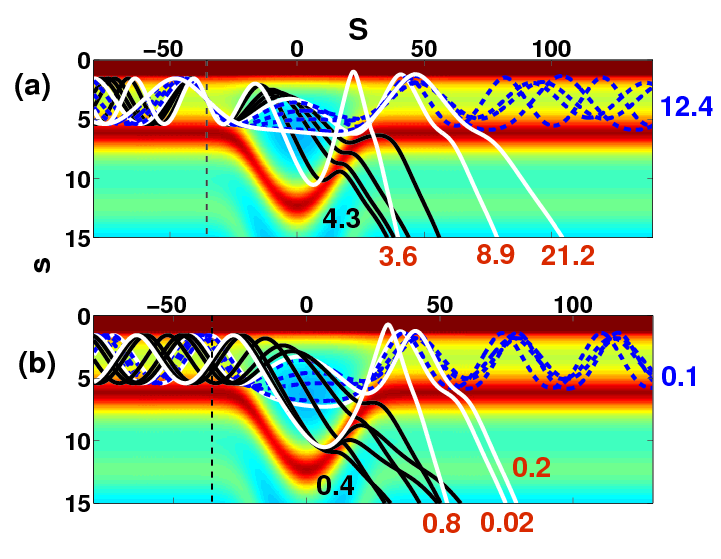}
\end{center}
\caption{\label{2traj} (color online) (a) Trajectories with different CM kinetic energies $T_S$ for $s_{0L} \approx1.56$ and $\frac{t}{T}=0.22$. 
The numerical values presented in the diagram correspond to the values of $T_S$ for the different kinds of trajectories.
(b) Trajectories with different phases of relative oscillation $\frac{t}{T}$ for $T_S=8.88$ and $s_{0L} \approx 1.56$.
The numerical values presented in the diagram correspond to the values of $\frac{t}{T}$ for the different kinds of trajectories.
In both cases the dissociative trajectories of type $A$ are shown with solid white lines, while those of type $B$ are represented with solid black lines. 
The finally bound trajectories are presented with
blue dashed lines. The vertical black dashed line indicates the position $S_{hL}$ at which the hump starts (by definition).}
\end{figure}

\textit{Trajectories.} As we have seen throughout this section the  initially bound trajectories are divided into  two different   categories: those which remain finally,
i.e. after scattering, bound and those which are led to dissociation.
 It is evident that the finally dissociated trajectories can be  further classified  into those that dissociate after reaching the uniform region (type $A$)
and the ones  that dissociate within the hump (type $B$). Since they reach the homogeneous domain (Fig. \ref{2traj}),  the dissociated trajectories of type $A$ have more 
features in common with the
 bound ones and this 
is the reason why they always occur close to the transition points. States of type $B$, on the other hand are fundamentally distinct (Fig. \ref{2traj}) and 
 occur only in the middle of the dissociation regions
(Fig.  \ref{scat1we} (c)). The main difference of $A$ and $B$ trajectories is imprinted in the energy transfer. Type $A$ trajectories  have always final relative energy greater 
than the potential barrier
and overall can be thought as cases where a substantial amount of energy has been transfered  from the CM to the relative motion.
 However, the trajectories of type $B$ pass
the potential barrier with their total amount of energy $E$. Since they remain in the inhomogeneous regime for some time after dissociation, a redistribution of energy   between the CM and the relative motion, is still possible leading to a sequence of loss and gain of relative energy. Therefore, their 
final relative energy $E_s$ can be lower than the height of the potential barrier (Fig.  \ref{scat1we} (c)).

\begin{figure}[htbp]
\begin{center}
\includegraphics[width=8.6cm]{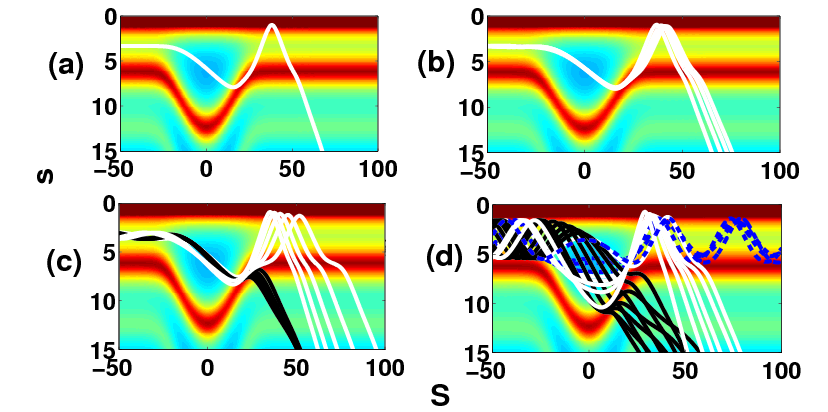}
\end{center}
\caption{\label{4traj} (color online) Trajectories for various phases of relative oscillation $\frac{t}{T}$ for $T_S=8.88$ and: (a) $s_{0L}=s_{min}=3.34$
(b) $s_{0L}=3.26$, (c)  $s_{0L}=2.53$, (d)  $s_{0L}=1.56$. Solid white lines stand  for finally dissociated states of type $A$, solid black lines for type $B$,  and
dashed blue lines for finally bound states.}
\end{figure}

 Figures \ref{2traj}, 
\ref{4traj}  specify the above line of arguments and identify in particular the different types of trajectories.  For a constant 
value of $s_{0L}$, sufficiently away from the potential minimum, one can induce transitions of the form: 
\[ \it{A} \rightarrow \it{B}  \rightarrow  \it{A} \rightarrow  \rm{bound~states} \rightarrow \it{A}, \]
by varying either the CM kinetic energy (increasing $T_S$ (Fig. \ref{2traj} (a))), or the phase of the relative oscillation (decreasing $\frac{t}{T}$, Fig. \ref{2traj} (b)). 

Figure  \ref{4traj} provides us with the complementary information of how the change of the initial relative energy $E_s$, imprinted in $s_{0L}$, affects the evolution of the trajectories.
The trajectories presented for each $s_{0L}$ have the same $T_S$ and different phases $\frac{t}{T}$. For $s_{0L}=s_{min}$ (Fig.  \ref{4traj} (a)), a case familiar from Subsec. A,  the
fictitious particle
does not oscillate and thus the trajectories are independent of the phase. For the value of $T_S$ chosen here, this set of trajectories constitutes a single dissociated state of
 type $A$. Increasing the relative energy (decreasing $s_{0L}$), the trajectories for various phases start to  separate, but still their type remains the same (Fig.  \ref{4traj} (b)).
A further increment of $E_s$ (Fig.  \ref{4traj} (c)) has as a result the formation of dissociative states of type $B$  for certain values of phases. Finally, for $s_{0L}$
sufficiently close to the potential barrier (Fig.  \ref{4traj} (d)), the amplitude of the relative oscillation increases dramatically allowing for the emergence of all the three types of 
trajectories, including finally bound states.

\textit{Energy transfer.} Throughout this section we have come across intriguing effects originating from the coupling between the CM and the relative coordinate. The key ingredient 
allowing for these effects is the energy transfer between the two degrees of freedom inside the  inhomogeneous region. Let us therefore point out some basic features of the energy exchange process. To do so, we consider the change of the kinetic energy of the CM $T_S$. 
The latter is zero in the uniform regime of the helix. From the equations of motion for the Hamiltonian (\ref{hmlt2})
 we obtain $\dot{T}_S=-\dot{S}\pd{V}{S}$, 
where $V=V(S,s)$. Although this relation for $\dot{T}_S$ refers to certain time evolving trajectories we find it instructive to analyze its contour plot
 for a certain value
of $\dot{S}=1$. 

\begin{figure}[htbp]
\begin{center}
\includegraphics[width=8.6cm]{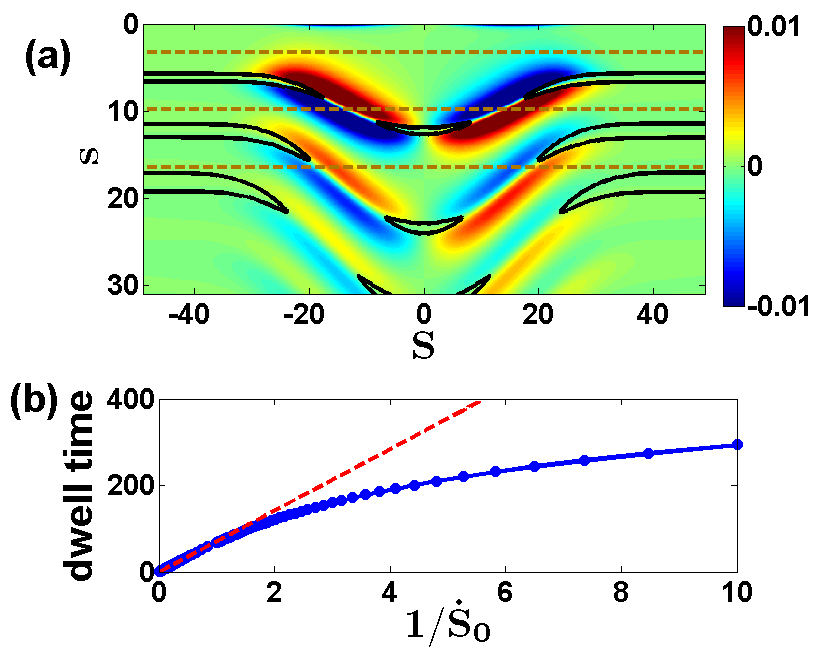}
\end{center}
\caption{\label{dotks} (color online)  (a) The rate of change of the kinetic CM energy $\dot{T}_S$ as a function of $S,s$ for $\dot{S}=1$.
 The solid black lines depict the position of the maxima of the three potential barriers,
while the dashed brown lines indicate the position of the minima of the three potential wells in the uniform regime (see also Fig. \ref{hpot}).
(b) The dwell time  as a function of the initial CM velocity $\dot{S}_0$ for the particles starting at the minimum of the first potential well with zero relative velocity 
(blue line with circles). For small $\dot{S}_0$ the deviation from a motion with  constant $\dot{S}=\dot{S}_0$ dictated by the red dashed line, is evident.}
\end{figure}

We observe in Fig. \ref{dotks} (a), that for a constant value of the CM velocity, the  rate of change of $T_S$ is non-zero only in
the inhomogeneous regime as expected.
Moreover it is  antisymmetric with respect to  the center of the hump $S=0$, meaning that if at $(-S,s)$ the particle gains $T_S$, it loses at $(S,s)$. 
Therefore, almost symmetric trajectories $(S \rightarrow -S)$, as those 
for very large or very small initial CM velocity $\dot{S}_0$, will have finally almost zero energy transfer. However, since $T_S$ - moving from a positive $\dot{T}_S$
value to a negative one - reaches a maximum for these 
 trajectories inside the inhomogeneity (at $S \approx 0$), 
the average kinetic energy of the  CM motion inside the hump will be  larger than that in the homogeneous regime.
This in turn leads to a larger  effective $\dot{S}$ within the inhomogeneity and a smaller dwell time (defined as the time interval during which the fictitious particle
 moves from $-S_{h}$ to $S_{h}$)
than the one expected by $\dot{S}_0$.  This effect
is evident  (Fig. \ref{dotks} (b)) only in the case of small $\dot{S}_0$, where even a slight increment in the velocity affects substantially the value of the dwell time.

 The greatest amount of energy can be gained or lost when the
fictitious  particle passes deep in the potential well, close to the potential barrier since  the gradient $\pd{V}{S}$ acquires there its largest values (Figs. \ref{hpot}, \ref{dotks} (a)).
 For $\dot{S}>0$ which is
always true for particles passing from the left homogeneous regime to the right one, the CM gains kinetic energy while being in the first potential well
in the region $S<0$, and it loses for  $S>0$. This causes highly asymmetric trajectories as some with $\dot{S}_0>1$ (Figs. \ref{2traj}, \ref{4traj}) overall to lose  an amount of CM kinetic energy, 
which after reaching the uniform regime appears as a gain in the total relative energy. Trajectories that dissociate within the hump (type $B$),
 after crossing the top of the potential
 barrier for $S>0$, regain CM kinetic energy $T_S$ (Figs. \ref{hpot}, \ref{dotks} (a)), but since they continue moving at lower values of the potential 
this does not always result in lower final values of total relative energy. Therefore, we may conclude that all the dissociative trajectories with
a lowered final $E_s$ belong to type $B$, but not vice-versa.


Overall, it is evident that the energy transfer consists of subsequent losses and gains of $T_S$ and $T_s$ induced by the variations of the potential $V(S,s)$ inside the inhomogeneous region, leading to a final asymptotic effective gain or loss.

\begin{center}
 { \bf{V. PHASE SPACE ANALYSIS}} 
\end{center}

We explore now  the structure
of the underlying phase space of the scattering process and in particular of bound states in the inhomogeneity of the helix. Since the three potential wells display similar characteristics, with the first one allowing for more  variations in energy since it is the deepest one,
it will be the only one we consider here. For Hamiltonian systems with two degrees of freedom the standard tool for such an analysis is the Poincar\'{e} surface
of section (PSOS), taking  advantage of the conservation of energy. Here, we will choose $S=0~(P>0)$ as the intersection through the energy shell. We note that the PSOS shown below
report only the bound state trajectories. 

\begin{figure}[htbp]
\begin{center}
\includegraphics[width=8.6cm]{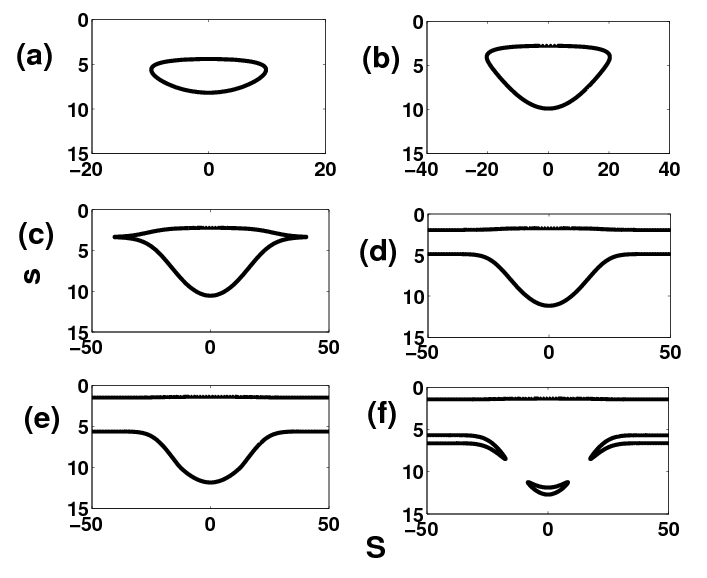}
\end{center}
\caption{\label{czva} (color online)  EPLs of the first potential well for representative total energies: (a) $E=0.28$, (b) $E=0.39$, (c) $E=0.476$, (d) $E=0.6$, (e) $E=0.744$
and (f) $E=0.76$. 
}
\end{figure}

Let us inspect the regions of bound and unbound motion within  the inhomogeneous helix. 
The only part of the potential landscape $V(S,s)$ that can support
bounded motion is that of the inhomogeneity, i.e. inside the hump, in the neighborhood of $S \approx 0$. 
Since the potential wells possess finite barriers, it is evident that for energies beyond a certain amount the fictitious particle  can escape to infinity concerning either
the CM $S$ or the relative coordinate $s$, leading to dissociation.
 
 For the first potential well this fact is clearly depicted (Fig. \ref{czva}) through the equipotential lines (EPLs).
 For $E \leq E_{c_1}=0.476$, with $E_{c_1}$ being the energy of the minimum of the first potential well in the uniform domain, the EPLs  are closed both 
in $S$- and $s$-direction (Fig. \ref{czva} (a)-(c)) leading to exclusively  bounded motion inside the hump. Figure \ref{czva} (c) presents the critical case,
 a fact that is reflected in the substantial elongation  of the
 wings of 
the EPL. A further increment of  the energy  leads to EPLs extending to $\abs{S} \rightarrow \infty$ (Figs. \ref{czva} (d)-(f))
which allows for escaping trajectories from the center of the hump to the homogeneous regime of the helix. This holds until the second critical
value  $E_{c_2}=0.744$  (Fig. \ref{czva} (e)) is reached. From then on, two additional openings are formed in the EPL inside the central region of the hump (Fig. \ref{czva} (f))
allowing also for  escapes in the relative coordinate.  For energies larger than the maximum value of the first potential well $V_{max}=0.81$ of course, the particles' motion is in 
principle unbounded.

\begin{figure}[htbp]
\begin{center}
\includegraphics[width=8.6cm]{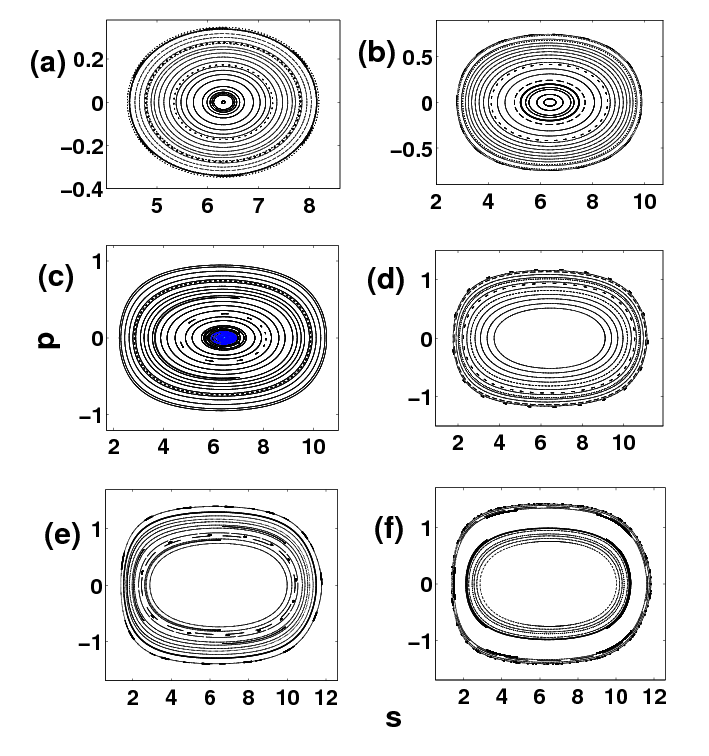}
\end{center}
\caption{\label{lsfs} (color online) Poincar\'{e} surfaces of section for representative total energies: (a) $E=0.28$, (b) $E=0.39$, (c) $E=0.476$, (d) $E=0.6$, (e) $E=0.744$
and (f) $E=0.76$. The blue dots denote chaotic trajectories. 
}
\end{figure}

Figure \ref{lsfs} shows PSOS for different energies. For $E<E_{c_1}$ (Fig. \ref {lsfs} (a), (b)) we observe an elliptic island. Close to the first transition
point $E=E_{c_1}$ (Fig. \ref {lsfs} (c)) it develops in its inner region, i.e. for small $p$ and $s$ close to the absolute minimum of the first potential well, 
a chaotic portion. A further increase of the energy leads to escaping trajectories which is evident in Fig. \ref{lsfs} (d)  where a large part of the inner region of the surface of
 section (empty region) belongs to escaping trajectories  $(\abs{S} \rightarrow \infty)$ through the respective 
openings of  the EPL (Fig. \ref {czva} (d)). As the energy approaches its second critical value $E_{c_2}$ (Fig. \ref {lsfs} (e)) the basin of escape becomes larger and
 finally for $E> E_{c_2}$  (Fig. \ref {lsfs} (f)), a second area  of the surface of section empties, this time in the center of the region of bounded motion. 
This corresponds to trajectories that escape in the $s$-direction (dissociation) through the two additional openings on the lower side of the corresponding EPL (Fig. \ref {czva} (f)). 

\begin{figure}[htbp]
\begin{center}
\includegraphics[width=8.6cm]{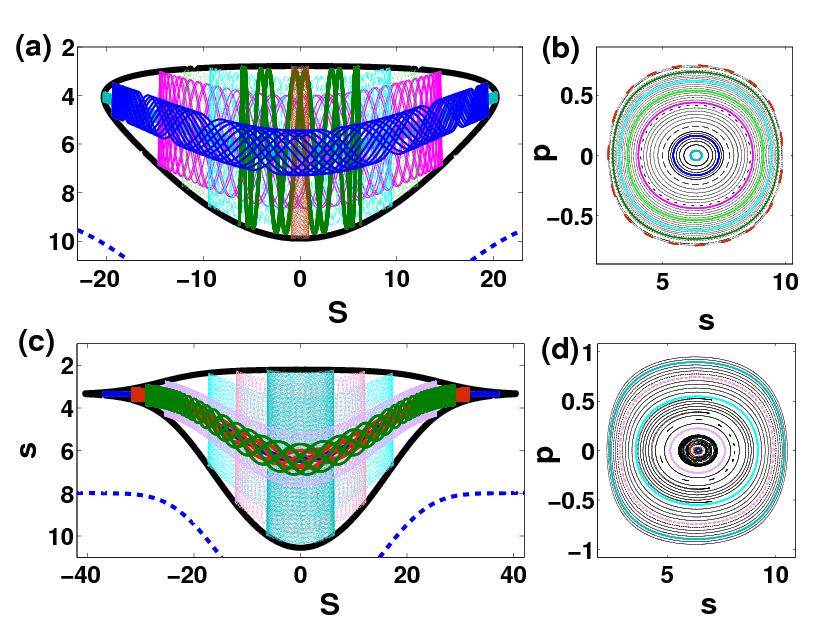}
\end{center}
\caption{\label{lstr1} (color online) Selected trajectories (a),(c) and respective PSOS (b), (d) for the energies: (a),(b) $E=0.39$, (c),(d) $E=0.4761$. The color
of the trajectories corresponds to their position in the PSOS. In (a),(c) the EPLs are also depicted  as in Fig. \ref{czva}.
}

\end{figure}

\begin{figure}[htbp]
\begin{center}
\includegraphics[width=8.6cm]{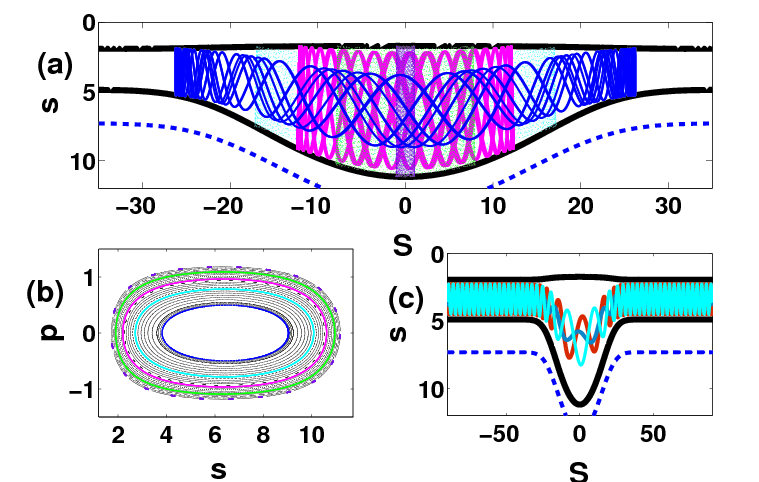}
\end{center}
\caption{\label{lstr2} (color online) (a) Selected bound trajectories, (b) PSOS and (c) escaping trajectories in the CM $S$ coordinate for the energy $E=0.6$.
}
\end{figure}

\begin{figure}[htbp]
\begin{center}
\includegraphics[width=8.6cm]{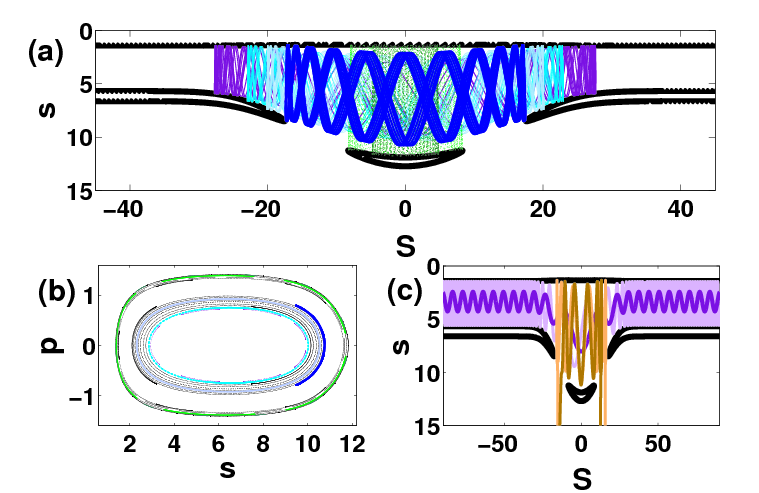}
\end{center}
\caption{\label{lstr3} (color online) (a) Selected bound trajectories, (b) PSOS and (c) escaping trajectories both in the CM $S$ (bound states) and in the relative 
 coordinate $s$ (free states) for energy $E=0.76$.
}
\end{figure}

Figures \ref{lstr1}, \ref{lstr2} and \ref{lstr3} provide further details enriching the above analysis in terms of specific trajectories.  First we remark
 that the turning points of the
trajectories both in $s$- and $S$-direction, lie on the EPLs of the respective energy. Trajectories  with a larger $S$ elongation and a
 reduced $s$-amplitude follow obviously the bending of the potential landscape and are the ones that map to the inner region of the Poincar\'{e} surface of section. 
As we move to the outer region of the surface of section the amplitude in $s$ increases with a respective decrease of the amplitude in the $S$-direction.
 For energies $E_{c_1}<E<E_{c_2}$
trajectories can escape from the left and right openings of the respective EPL (Fig. \ref{lstr2} (c)). In order to be bound, the trajectories should have an amplitude of the relative 
motion exceeding the width of the openings of the EPL (Fig. \ref{lstr2} (a)). 
This is the reason why for the PSOS the  regime of bounded motion is located in the outer part, 
with the inner one being empty and corresponding to  escaping trajectories (Fig. \ref{lstr2} (b)).

A more detailed nonlinear dynamical analysis would most probably reveal two unstable periodic orbits that provide the connection
between the bounded and escaping motion. To explore this in detail goes however beyond the scope of the present
 work which has its emphasis on the main phenomena appearing in the
helical dynamics investigated here for the first time. 

\begin{figure}[htbp]
\begin{center}
\includegraphics[width=8.6cm]{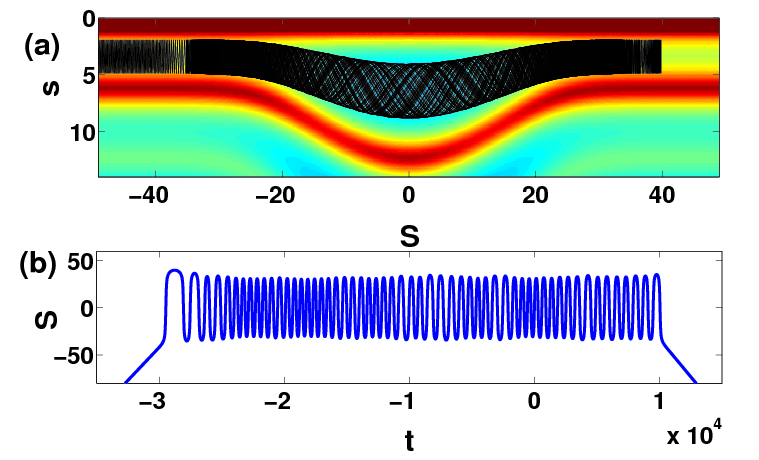}
\end{center}
\caption{\label{rest1} (color online) A resonant trajectory for $E=0.6$ and initial conditions $S=0, ~s=5.134,~\dot{S}=0.549, ~\dot{s}=-0.387$: (a) plot of the trajectory on the potential 
landscape,
(b) time evolution of the CM coordinate $S$ of  the trajectory.
}
\end{figure}

Bridging between bound and unbound there are ''resonant''  trajectories, i.e. trajectories that remain within the hump performing oscillatory CM and  
relative motion, for a large time interval and finally escaping to the homogeneous asymptotic region. They have typical initial conditions  in the empty region
of the PSOS (Fig. \ref{lstr2} (b)), close to the innermost bound trajectory. Such a trajectory is presented in Fig. \ref{rest1}.  It escapes to the left opening of the potential well ($S<0$)
both in forward and in backward propagation time, i.e. it is reflected at the helical hump.   

A more complex structure  of the PSOS is encountered for $E_{c_2}<E<V_{max}$. In such a case, as we have remarked earlier, there are four openings of the EPL.
 Subsequently, four unstable periodic  orbits exist and thus four possibilities for escape symmetric with respect to $S=0$ (Fig. \ref{lstr3} (c)): two in the CM coordinate direction 
($S$-openings) as before and
two in the relative one ($s$-openings). The escape in the $S$-direction (small $s$, large $S$, i.e. $s_1 s_2>0$) both in forward
 and in backward propagation time, corresponds to a bound pair
of particles that after scattering within the hump remains bound, a case that has been referred in the previous section as a bound
state. On the contrary, escapes in $s$-direction (large $s$, small $S$)
correspond to free particles that come from opposite sides of the helical trap ($s_1 s_2<0$), scatter within the hump and return backwards in opposite directions. They thus account for conventional
 scattering of free particles and are different from the peculiar dissociative states we have observed so far.

Between the regimes of escapes, there are two distinct regions of bounded trajectories (blue and green, Fig. \ref{lstr3} (a)).
The first of them consists of trajectories supported by the side parts of the EPL.
 Their relative oscillation $s$-amplitude is bounded both from below, by the width of the $S$-opening in the uniform domain, and from above by its edge width inside the 
inhomogeneous region.
The second  consists of bound states with somewhat larger relative $s$-amplitude dictated by the width of the central part of the EPL. 
These considerations, give  rise to the  two-ring
structure of the PSOS (Fig. \ref{lstr3} (b)).

\begin{center}
 { \bf{VI. SUMMARY AND OUTLOOK}} 
\end{center}

We have investigated the classical scattering of Coulomb interacting particles in a helix. First, we have proven that 
for two particles interacting via a potential that depends exclusively on their Euclidean distance, a separation of the CM,
 leading to a conservation of the total momentum 
is provided if and only if the confining curve is a homogeneous helix. Having this result in mind, we investigated the scattering of charged particles from a local inhomogeneity of the helix. In such a
system the coupling between the CM and the relative degrees of freedom induces intriguing effects. The most important of them is the dissociation of initially bound states 
of the two repulsively interacting charged particles through scattering. Due to the time reversal symmetry imprinted in the equations of motion, this leads to the conclusion that initially unbound charged particles can become bound when
 scattered, a fact counterintuitive regarding especially the repulsive character of the interaction.

  The underlying mechanism for such a behaviour was found to be the effective energy transfer 
between the relative and the CM motion occurring due to their coupling. It has been pointed out, nevertheless, that this transfer does not take place in a single step,
but it is the final result of the continuous energy redistribution in the whole time interval in which the particles remain inside the hump. For this reason, the outcome of the scattering of
initially bounded charged particles in terms of finally bound and dissociated states, depends in a rather complex way on the initial conditions. The dissociation regimes depending on these 
condition have been identified and analyzed in detail. The most important parameter is shown to be  the value of the initial  kinetic energy of the CM $T_S$. In most cases for
 very small or very large values of $T_S$  the particles remain bound after the scattering, with dissociation occurring only in the intermediate regime. This is attributed  to the
 little amount of energy available for transfer and to the small dwell time in the hump that prevents a strong coupling of the relative
and CM degrees of freedom respectively.

Our analysis  has been completed with an exploration of the phase space structure of the deepest potential well that can support bound states.  Regimes of bound regular motion
inside the hump, as well as regimes of escapes were identified by varying the total energy. This exploration provided us with  bound states localized inside the hump, as well as 
with ''resonant trajectories``.  

Further studies could be dedicated to a more detailed investigation of the phase space  searching for stable and unstable periodic orbits and their asymptotic curves, a fact that would
 allow a 
rigorous and quantitative analysis of the escape procedure. A promising direction is the study of many-body systems which are expected to exhibit an intriguing dynamics
 as well as leading to exceptional
transition phenomena. 
\newpage
\begin{center}
 { \bf{APPENDIX: CONDITIONS ON THE CURVED 1D-MANIFOLD FOR THE SEPARATION OF THE CENTER OF MASS FROM THE RELATIVE MOTION}} 
\end{center}
\textbf{Proposition:} \textit{The condition:
\begin{equation}
 \partial_{s_1} R(s_1,s_2)+\partial_{s_2} R(s_1,s_2)=0, ~\forall s_1,s_2 \in \mathbb{R}, \label{con1}
\end{equation}
 where $R(s_1,s_2) = \abs{\vec{r}(s_1)-\vec{r}(s_2)}$ holds for a smooth, regular curve $\vec{r}(s)$ in arc length parametrization if and only if the curve is a homogeneous helix (including the degenerate cases of a circle or a straight line). }
\newline
\\
\textbf{Proof:} 
''$\Leftarrow$``: By the discussion in section II, condition (\ref{con1}) is equivalent to separability of CM and relative motion, which has been demonstrated to hold for a homogeneous helix (see discussion in the main text).

''$\Rightarrow$``: Assume that condition (\ref{con1}) holds. The outline of the proof is as follows. We show that by virtue of Eq. (\ref{con1}), for each $x \in \mathbb{R}$ the map
\begin{equation}
 F_x: \mathcal{W} \mapsto \mathcal{W},~ F_x(\vec{r}(s))=\vec{r}(s+x). \label{fis1}
\end{equation}
is an isometry (i.e. it preserves distances) on the submanifold $\mathcal{W} \subseteq \mathbb{R}^3$, which is defined as the image of the curve $\vec r$. We extend this family of isometries to a family of isometries $\left\{\Gamma_x\right\}_{x \in \mathbb R}$ from all of $\mathbb R^3$ into itself, with the property that the restriction $\Gamma_x \big|_\mathcal{W} = F_x$. The isometries of $\mathbb{R}^3$ form the group of Euclidean moves $E(3)$. Since the $\Gamma_x$ are continuously deformed to the identity map for $x \rightarrow 0$, they belong to the identity component of $E(3)$, i.e. to $SE(3)$. Thus by the classification theorem for Euclidean moves \cite{Tarrida} each $\Gamma_x$ is a screw operation or a degenerate case thereof, i.e. a pure rotation or translation. Furthermore, not all $\Gamma_x$ can be the identity on $\mathbb{R}^3$, since then the curve $\vec r$ would have the property $\vec r(s+x) = \vec r(s)$ for all $x$ and would degenerate to a point. Thus, there is at least one non-trivial screw operation $\Gamma_x$ that maps the curve $\mathcal W$ to itself, so $\mathcal W$, the image of $\vec r(s)$, must be a (homogeneous) helix, which proves the proposition.

\medskip
We now proceed to the detailed proof and first show that, given Eq. (\ref{con1}), the map $F_x$ as defined in (\ref{fis1}) is an isometry on $\mathcal W$. 
First we prove that there is a function $\chi$ such that $R(s_1,s_2)=\chi(s_1 - s_2)$. To see this, introduce new variables $\xi_- := s_1-s_2,~\xi_+:= s_1+s_2$
and a function $\chi$ with the property $\chi(\xi_+,\xi_-)=R(s_1,s_2)$. Then condition (\ref{con1}) yields $\partial \chi / \partial \xi_+=0$, leading to $\chi(\xi_+,\xi_-)=\chi(\xi_-)$ or
\begin{equation*}
 R(s_1,s_2)=\chi(s_1-s_2) \label{con2}.
\end{equation*}
This, in turn, immediately implies $R(s_1,s_2) = R(s_1 + x, s_2 + x)$, or 
\begin{equation*}
 \abs{\vec{r}(s_1)-\vec{r}(s_2)}=\abs{\vec{r}(s_1+x)-\vec{r}(s_2+x)} \label{con3}
\end{equation*}
for all $x,s_1,s_2 \in \mathbb{R}$, showing that indeed $F_x$ as defined above is an isometry on $\mathcal W$. 

We now assume without loss of generality that $\vec 0 \in \mathcal W$ and proceed to show that for all $x$ the map \[\tilde{F}_x: \mathcal{W} \mapsto \mathbb{R}^3,~\tilde{F}_x (\vec{r}):=F_x(\vec{r})-F_x(\vec{0})\]
has the following properties:\\
(i) $\abs{\tilde{F}_x(\vec r)} = \abs{\vec r} \,\,\,\,\,\, \forall \vec r \in \mathcal W$.\\
(ii)  $\sigma\left( \tilde{F}_x(\vec r_1), \tilde{F}_x(\vec r_2) \right) = \sigma\left( \vec r_1, \vec r_2 \right) \,\,\,\,\, \forall \vec r_1, \vec r_2 \in \mathcal W$ .\\
(iii) $\tilde{F}\left(\alpha_1 \vec r_1 + \alpha_2 \vec r_2 \right) = \alpha_1 \tilde{F}_x  \left(\vec r_1 \right) + \alpha_2 \tilde{F}_x  \left(\vec r_2 \right)$ \\
\phantom{...........}$ \forall \vec r_1, \vec r_2 \in \mathcal W, \alpha_1, \alpha_2 \in \mathbb{R} \text{ s.t. } \alpha_1 \vec r_1 + \alpha_2 \vec r_2 \in \mathcal W$.\\
Here $\sigma$ denotes the Euclidean scalar product. (i) immediately follows from $F_x$ being an isometry on $\mathcal W$.
(ii) follows from (i) and $F_x$ being an isometry, since $\forall \vec{r}_1,\vec{r}_2 \in \mathcal{W}$:
\begin{eqnarray*}
 &&2 \sigma(\tilde{F}_x(\vec{r}_1),\tilde{F}_x(\vec{r}_2))\\
&=& \abs{\tilde{F}_x(\vec{r}_1)}^2 + \abs{\tilde{F}_x(\vec{r}_2)}^2-\abs{\tilde{F}_x(\vec{r}_1)-\tilde{F}_x(\vec{r}_2)}^2 \nonumber \\
               &=& \abs{\vec{r}_1}^2+ \abs{\vec{r}_2}^2-\abs{\vec{r}_1-\vec{r}_2}^2=  2 \sigma(\vec{r}_1,\vec{r}_2). \label{dotpr1}
\end{eqnarray*}
Finally, using (i) and (ii) it is easily shown that
\begin{eqnarray*}
\abs{\tilde{F}_x (\alpha_1 \vec{r}_1+\alpha_2 \vec{r}_2)-\alpha_1 \tilde{F}_x(\vec{r}_1)-\alpha_2 \tilde{F}_x(\vec{r}_2)}^2 \nonumber \\
=\abs{ (\alpha_1 \vec{r}_1+\alpha_2 \vec{r}_2)-\alpha_1 \vec{r}_1-\alpha_2 \vec{r}_2}^2=0 \nonumber,
\end{eqnarray*}
which proves (iii).

Now we are in the position to construct the extended isometries $\Gamma_x$. Let us first assume that the curve $\vec r$ does not entirely lie in a plane. 
Then we can form a basis of $\mathbb{R}^3$ with three linearly independent
vectors $\vec{w}_i \in \mathcal{W}$. Hence for each $\vec{x} \in \mathbb{R}^3$ there exists a unique expansion $\vec{x}= \sum_{i=1}^3 \alpha_i \vec{w}_i,~\alpha_i \in \mathbb{R}$.
For any such $\vec{x}$ we define 
\begin{equation*}
 \Gamma_x(\vec{x}):=F_x(\vec{0})+\sum_{i=1}^3 \alpha_i \tilde{F}_x (\vec{w}_i). \label{apeq2}
\end{equation*}
Evidently, for the special case of $\vec{r} = \sum_{i=1}^3\gamma_i \vec{w}_i \in \mathcal{W}$:
\begin{eqnarray*}
 \Gamma_x(\vec{r})&=&F_x(\vec{0})+\sum_{i=1}^3 \gamma_i \tilde{F}_x (\vec{w}_i)=F_x(\vec{0})+ \tilde{F}_x \left(\sum_{i=1}^3 \gamma_i \vec{w}_i\right) \nonumber \\
&=& F_x(\vec{0})+\tilde{F}_x(\vec{r})=F_x(\vec{r}), \label{apeq3}
\end{eqnarray*}
due to property (iii) of $\tilde{F}_x$, such that indeed the restriction $\Gamma_x \rvert_\mathcal{W} =F_x$.
Now, using property (ii) of $\tilde{F}_x$, it is straightforward to show that for any $\vec x$ and $\vec{y}=\sum_{i=1}^3 \beta_i \vec{w}_i \in \mathbb{R}^3$:
\begin{eqnarray*}
 &&\abs{\Gamma_x(\vec{x})-\Gamma_x(\vec{y})}^2=\abs{\sum_{i=1}^3 (\alpha_i-\beta_i) \tilde{F}_x (\vec{w}_i)}^2 \nonumber \\
&=&\abs{\sum_{i=1}^3 (\alpha_i-\beta_i)  \vec{w}_i}^2 =\abs{\vec{x}-\vec{y}}^2
\end{eqnarray*}
which proves that $\Gamma_x$ is an isometry of $\mathbb{R}^3$.

Finally, we address the special case of a planar curve. Then either $\mathcal W$ is a straight line, in which case there is nothing to prove, since this is a degenerate case of a helix.
Otherwise, we pick two linearly independent vectors $\vec{w}_i \in \mathcal W$ and a third vector
$\vec{k}_3$ perpendicular to $\vec{w}_1,\vec{w}_2$. Since $F_x$ maps $\mathcal{W}$ to itself it is clear that $\sigma(\tilde{F}_x(\vec{w}_i),\vec{k}_3)=0$ as well. 
For any vector $\vec{x} \in \mathbb{R}^3$ a representation $\vec{x}= \alpha_1 \vec{w}_1+ \alpha_2 \vec{w}_2 + \alpha_3 \vec{k}_3$
is possible and we define:
\begin{equation*}
 \Gamma_x(\vec{x})=F_x(\vec{0})+\alpha_1 \tilde{F}_x(\vec{w}_1)+\alpha_2 \tilde{F}_x(\vec{w}_2)+\alpha_3 \vec{k}_3, \label{apeq4}
\end{equation*}
which for $\vec{x} \in \mathcal{W}$ (implying $\alpha_3=0$) using (iii) again leads to $\Gamma_x \rvert_\mathcal{W} =F_x$.
Furthermore, using the orthogonality of $\vec{k}_3$ to 
$\vec{w}_i, \tilde{F}_x(\vec{w}_i)$ as well as (ii), it follows for any $\vec{y}= \sum_{i=1}^2 \beta_i \vec{w}_i+ \beta_3 \vec{k}_3$ that:
 \begin{eqnarray*}
 &&\abs{\Gamma_x(\vec{x})-\Gamma_x(\vec{y})}^2 =\abs{\sum_{i=1}^2 (\alpha_i-\beta_i) \tilde{F}_x (\vec{w}_i)+(\alpha_3-\beta_3) \vec{k}_3}^2 \nonumber \\
&=&\abs{\sum_{i=1}^2 (\alpha_i-\beta_i)  \vec{w}_i+(\alpha_3-\beta_3) \vec{k}_3}^2 = \abs{\vec{x}-\vec{y}}^2.
\end{eqnarray*}
Therefore for this case, too, one can construct an isometry $\Gamma_x$ of $\mathbb{R}^3$ which extends $F_x$. 
Evidently, in both cases tuning $x \rightarrow 0$ one can continuously transform the $\Gamma_x$ to $\Gamma_{x=0} = \text{id}_{\mathbb{R}^3}$, such that all $\Gamma_x$ lie in $SE(3)$. 

\begin{center}
 { \bf{ACKNOWLEDGEMENTS}} 
\end{center}
We thank M. J\"{u}ngling for many helpful discussions and his support concerning the theoretical conceptual aspects of this work.
A. Z. thanks the International Max Planck Research School for Ultrafast Imaging and Structural Dynamics for a PhD scholarship.
 J. S. and S. K. gratefully acknowledge funding by the Studienstiftung des deutschen Volkes.


\begin{thebibliography}{99}
\bibitem{footnote} The property of having either a closed or infinitely extended curve can be formalised in terms of the arc length: For any arc length $s$, 
$\vec{r}(s)$ must be well defined, i.e. a point of the space curve under consideration.
\end{thebibliography}
\end{document}